\RequirePackage{lineno}
\documentclass[twocolumn,aps,tightenlinefs,superscriptaddress,preprintnumbers]{revtex4}

\usepackage{graphicx,color,dcolumn,booktabs,bm}
\usepackage{longtable,lscape}
\usepackage{amsmath}
\usepackage{indentfirst}
\usepackage{epsfig}
\usepackage{feynmf}   
\usepackage{epstopdf}   
\usepackage{slashed}  
\usepackage{color}
\usepackage{multirow}
\usepackage{ulem}
\usepackage{verbatim}
\usepackage[colorlinks,linkcolor=red,anchorcolor=green,citecolor=blue]{hyperref}
\usepackage{mathrsfs}
\usepackage{rotating}
\usepackage{threeparttable}
\usepackage{lineno}
\usepackage{subfigure}

\graphicspath{{ps}}

\newcommand{\psip}{\psi'}
\newcommand{\etacp}{\eta_c(2S)}

\newcommand{\tripipi}{3(\pi^+\pi^-)}

\newcommand{\BF}{\mathcal{B}}
\newcommand{\ra}{\rightarrow}

\newcommand{\etacpD}{\eta_c(2S)\rightarrow3(\pi^+\pi^-)}

\begin{document}

\title{ \quad\\[0.1cm] \boldmath  Observation of $\etacp \ra \tripipi$ and measurements of $\chi_{cJ} \ra \tripipi$ in $\psi(3686)$ radiative transitions}

\author{
M.~Ablikim$^{1}$, M.~N.~Achasov$^{11,b}$, P.~Adlarson$^{70}$, M.~Albrecht$^{4}$, R.~Aliberti$^{31}$, A.~Amoroso$^{69A,69C}$, M.~R.~An$^{35}$, Q.~An$^{66,53}$, X.~H.~Bai$^{61}$, Y.~Bai$^{52}$
% ...
\\
\vspace{0.2cm}
(BESIII Collaboration)\\
\vspace{0.2cm} {\it
$^{1}$ Institute of High Energy Physics, Beijing 100049, People's Republic of China\\
% ...
$^{4}$ Bochum Ruhr-University, D-44780 Bochum, Germany\\
% ...
$^{11}$ G.I. Budker Institute of Nuclear Physics SB RAS (BINP), Novosibirsk 630090, Russia\\
% ...
$^{31}$ Johannes Gutenberg University of Mainz, Johann-Joachim-Becher-Weg 45, D-55099 Mainz, Germany\\
% ...
$^{35}$ Liaoning Normal University, Dalian 116029, People's Republic of China\\
% ...
$^{52}$ Southeast University, Nanjing 211100, People's Republic of China\\
$^{53}$ State Key Laboratory of Particle Detection and Electronics, Beijing 100049, Hefei 230026, People's Republic of China\\
% ...
$^{61}$ University of Jinan, Jinan 250022, People's Republic of China\\
% ...
$^{66}$ University of Science and Technology of China, Hefei 230026, People's Republic of China\\
% ...
$^{69}$ University of Turin and INFN, (A)University of Turin, I-10125, Turin, Italy; (B)University of Eastern Piedmont, I-15121, Alessandria, Italy; (C)INFN, I-10125, Turin, Italy\\
$^{70}$ Uppsala University, Box 516, SE-75120 Uppsala, Sweden\\
% ...
$^{b}$ Also at the Novosibirsk State University, Novosibirsk, 630090, Russia\\
% ...
}
}

\begin{abstract}
     The hadronic decay $\etacp \ra 3(\pi^+\pi^-)$ is observed with a
     statistical significance of 9.3 standard deviations using
     $(448.1\pm2.9)\times10^6$ $\psi(3686)$ events collected by the
     BESIII detector at the BEPCII collider. The measured mass and
     width of $\etacp$ are $(3643.4 \pm 2.3 ~(\rm stat.) \pm 4.4 ~(\rm
     syst.))$ MeV/$c^2$ and $(19.8 \pm 3.9~ (\rm stat.) \pm 3.1~ (\rm
     syst.))$ MeV, respectively, which are consistent with the world
     average values within two standard deviations.  The product
     branching fraction $\BF[\psi(3686)\to \gamma
       \eta_c(2S)]\times\BF[\eta_c(2S)\to3(\pi^+\pi^-)]$ is measured
     to be $(9.2 \pm 1.0~ (\rm stat.) \pm 1.2~ (\rm
          syst.))\times10^{-6}$. Using $\BF[\psi(3686)\to \gamma
       \eta_c(2S)]=(7.0^{+3.4}_{-2.5})\times10^{-4}$, we obtain
     $\BF[\etacp \ra 3(\pi^+\pi^-)] = (1.31 \pm 0.15 ~(\rm stat.) \pm
          0.17 ~(\rm syst.)~ (^{+0.64}_{-0.47}) ~(\rm extr))\times10^{-2}$,
     where the third uncertainty is from $\BF[\psi(3686) \ra \gamma
       \etacp]$. We also measure the $\chi_{cJ} \ra \tripipi$ ($J=0,
     1, 2$) decays via $\psip \ra \gamma \chi_{cJ}$ transitions. The
     branching fractions are $\BF[\chi_{c0} \ra \tripipi] =
     (2.080\pm0.006 ~(\rm stat.)\pm0.068 ~(\rm syst.))\times10^{-2}$,
     $\BF[\chi_{c1} \ra \tripipi] = (1.092\pm0.004 ~(\rm stat.)\pm0.035
     ~(\rm syst.))\times10^{-2}$, and $\BF[\chi_{c2} \ra \tripipi] =
     (1.565\pm0.005 ~(\rm stat.)\pm0.048 ~(\rm syst.))\times10^{-2}$.
\end{abstract}

\maketitle

\tighten
\lefthyphenmin=2
\righthyphenmin=2
\uchyph=0

\section{Introduction}

In recent years, remarkable experimental and theoretical progress has
been made in charmonium studies. All predicted charmonium states
below the open-charm threshold have been observed experimentally, and
to a large extent, the measured spectra agree with the theoretical
predictions based on Quantum Chromodynamics
(QCD)~\cite{qcd1,qcd2,qcd3} and QCD-inspired potential
models~\cite{qcd4,qcd5,qcd6}. However, there are still
problems or puzzles that need to be understood.

It is predicted that the ratio of branching fractions of $\psi(3686)$
and $J/\psi$ decaying into the same light hadron final states is
around 12\%, which was first proposed by Appelquist and Politzer using
perturbative QCD~\cite{bfr1}.  This is valid in most of the measured
hadronic channels~\cite{bfr2}, except for the decays into pseudoscalar
vector pairs and vector tensor pairs, which are suppressed by at least
an order of magnitude. 

The $\eta_c(2S)$ and $\eta_c(1S)$ are the spin-singlet partners of
$\psip$ ($\psi' \equiv \psi(3686)$) and $J/\psi$, but two different
ratios are calculated by the authors of Refs.~\cite{bfr4,bfr5}, suggesting two possibilities:

     \begin{equation}
     \nonumber
     \frac{\BF[\eta_c(2S)\ra hadrons]}{\BF[\eta_c(1S)\ra hadrons]} \simeq \frac{\BF[\psip \ra hadrons]}{\BF[J/\psi \ra hadrons]} \simeq 12\%,
     \end{equation} 

     or

     \begin{equation}
     \nonumber
     \frac{\BF[\eta_c(2S)\ra hadrons]}{\BF[\eta_c(1S)\ra hadrons]} \simeq 1,
     \end{equation}
respectively.  Using information on light hadronic final
states~\cite{pdg}, the authors of Ref.~\cite{bfr6} recently examined this
branching fraction ratio in several decay modes and found that the
experimental data are significantly different from both of the two
theoretical predictions, e.g., $\frac{\BF[\eta_c(2S)\ra \bar{K}K\pi]}{\BF[\eta_c(1S)\ra \bar{K}K\pi]}= 0.27^{+0.10}_{-0.07}$~\cite{bfr6}. 

The $\etacp$ is the first excited state of the pseudoscalar ground
state $\eta_c(1S)$, lying just below the mass of its vector
counterpart, $\psip$. It was first observed by the Belle Collaboration
in $B$ meson decay, $B^{\pm} \ra K^{\pm}\etacp$, in the exclusive
decay $\etacp \ra K_S^0K^{\pm}\pi^{\mp}$~\cite{Hist1}. This state was
hereafter confirmed by {\it BABAR}~\cite{Hist2}, CLEO~\cite{Hist3},
and Belle~\cite{Hist4} in the two-photon fusion process $\gamma\gamma \ra
\etacp \ra \bar{K}K\pi$, and by {\it BABAR}~\cite{Hist5} and
Belle~\cite{Hist6} in the double charmonium production process $e^+e^- \ra
J/\psi + c\bar{c}$. The magnetic dipole (M1) transition between
$\psip$ and $\etacp$ was first observed by BESIII, where $\etacp$ was
reconstructed with the $K\bar{K}\pi$ mode~\cite{ks0kpi}. Currently, only
seven decay modes of $\etacp$ have been observed experimentally, of which
the branching fractions of four have been measured with uncertainties
larger than 50\%~\cite{pdg}. The sum of the four decay modes is around
5\% of the total width of $\etacp$. Among the discovered decay modes
of $\eta_c(1S)$, the decay rate of $\eta_c(1S) \ra \tripipi$ is
relatively large, while the decay of $\etacp \ra \tripipi$ has not yet been
seen. CLEO searched for this mode in $\psip$ radiative decay
but no signal was observed~\cite{cleowork}.

The hadronic decays $\chi_{cJ} \ra \tripipi$ ($J=0,~1,~2$) were
successively measured by the MARK I collaboration in
1978~\cite{chicjto6pi1} and the BES collaboration in
1999~\cite{chicjto6pi2}. Since then, the results have not been
updated. In this work, we update the measurement of $\chi_{cJ} \ra
\tripipi$ with much higher statistics.

We present herein a study of the $\etacp \ra \tripipi$ and $\chi_{cJ}
\ra \tripipi$ in $\psip$ radiative transitions.
The measurement is based on a data sample corresponding to an
integrated luminosity of 668.55~$\rm pb^{-1}$
($(448.1\pm2.9)\times10^6$ $\psip$ events~\cite{psipNum}) produced at
the peak of the $\psip$ resonance and collected by the BESIII detector
at the BEPCII collider.  Additional datasets recorded at the
c.m.\ energies of 3.581 and 3.670 GeV with integrated luminosities of
85.7 and 84.7 $\rm pb^{-1}$, respectively, are used to determine the nonresonant
continuum background contributions.

\section{BESIII detector and Monte Carlo simulation}

The BESIII detector~\cite{bes3detec} records $e^+e^-$ collisions
provided by the BEPCII storage ring~\cite{bepc}, which operates with a
peak luminosity of $1\times10^{33}$ $\rm cm^{-2}s^{-1}$ in the
center-of-mass (c.m.) energy range from 2.00 to 4.95 GeV. The
cylindrical core of the BESIII detector covers 93\% of the full solid
angle and consists of a helium-based multilayer drift chamber (MDC), a
plastic scintillator time-of-flight system (TOF), and a CsI(Tl)
electromagnetic calorimeter (EMC), which are all enclosed in a
superconducting solenoidal magnet providing a 1.0 T magnetic
field. The solenoid is supported by an octagonal flux-return yoke with
resistive plate counter muon identification modules interleaved with
steel. The charged-particle momentum resolution at 1 GeV/$c$ is 0.5\%, and
the $dE/dx$ resolution is 6\% for the electrons from Bhabha scattering
at 1 GeV. The EMC measures photon energies with a resolution of 2.5\%
(5\%) at 1 GeV in the barrel (end cap) region. The time resolution of
the TOF barrel part is 68 ps, while that of the end cap part is 110 ps.

Simulated data samples produced with {\sc geant4}-based~\cite{geant4}
Monte Carlo (MC) software, which includes the geometric description of
the BESIII detector and the detector response, are used to determine
detection efficiencies and to estimate background contributions. The
simulation models the beam energy spread and initial state
radiation~(ISR) in the $e^+e^-$ annihilations with the generator {\sc
  kkmc}~\cite{kkmc}. The generic MC sample includes the production
of the $\psip$ resonance, the ISR production of the $J/\psi$, and the
continuum processes incorporated in {\sc kkmc}~\cite{kkmc}. The known
decay modes are generated with {\sc evtgen}~\cite{evtgen} using
branching fractions taken from the Particle Data Group~\cite{pdg}, and
the remaining unknown charmonium decays are modeled with {\sc
  lundcharm}~\cite{lundcharm}. Final state radiation~(FSR) from
charged final state particles is incorporated using {\sc
  photos}~\cite{photos}. The exclusive decays of $\psip \ra \gamma X$
are generated following the angular distribution of
$(1+\lambda\cos^2\theta)$, where $X$ refers to $\etacp$ or
$\chi_{cJ}$, $\theta$ is the polar angle of the radiative photon in
the rest frame of $\psip$, and the value of $\lambda$ is set to $1$ for
$\eta_c(2S)$ and to $1$, $-1/3$, $1/13$ for
$\chi_{cJ}~(J=0,~1,~2)$~\cite{chicj}, respectively. The $X \ra \tripipi$ decay is
generated uniformly in phase space (PHSP).  Additionally, two exclusive
MC samples, $\psip \ra \pi^0\tripipi$ and $\psip \ra (\gamma_{\rm
  FSR})\tripipi$, are generated according to PHSP to estimate
background contamination.

\section{Event selection}
We search for $\etacp$ in the exclusive decay $\psip \ra \gamma\etacp$
with $\etacp \ra \tripipi$ in events containing at least one radiative photon
and six charged tracks. Charged tracks detected in
the MDC are required to be within a polar angle ($\theta$) range of
$|\cos\theta|<0.93$, where $\theta$ is defined with respect to the
symmetry axis of the MDC ($z$-axis). For charged tracks, the distance
of closest approach to the interaction point must be less than
10\,cm along the $z$-axis, and less than 1\,cm in the transverse
plane. Charged-particle identification (PID) is based on the combined
information from the energy deposited in the MDC~(d$E$/d$x$) and the
flight time measured by the TOF, which are used to determine a
variable $\chi^2_{\rm PID}(h)$ for each track, where $h$ denotes the
pion, kaon, or proton hypothesis.

Photon candidates are identified using isolated showers in the EMC.  The
deposited energy of each shower must be larger than 25~MeV in both the
barrel region ($|\cos \theta|< 0.80$) and end cap region ($0.86 <|\cos
\theta|< 0.92$).  To suppress electronic noise and showers unrelated
to the event, the difference between the EMC time and the event start
time is required to be within (0, 700)\,ns.

Candidate events having exactly six charged tracks with net charge
zero and at least one candidate photon are retained. To improve the
mass resolution and suppress background, the total four-momentum
of the charged tracks and the photon candidate is constrained to the
initial $\psip$ four-momentum by a kinematic fit (4C fit). If there
is more than one photon candidate, the one resulting in the minimum $\chi^2$
from the 4C fit ($\chi^2_{\rm 4C}$) is selected as the radiative
photon. The background due to incorrect PID is suppressed by using
$\chi^2_{\rm tot} = \chi^2_{\rm 4C} + \sum_i \chi^2_{\rm PID}(\pi)$,
where $i$ runs over the six charged tracks. The candidate events
satisfying $\chi^2_{\rm tot} < 200$ and $\chi^2_{\rm 4C} < 44$ are
kept as the $\gamma\tripipi$ candidate events.

Background events from the $\psip \ra \pi^+\pi^-J/\psi$ process are
removed by a $J/\psi$ veto, which requires the recoil masses of
all $\pi^+\pi^-$ pairs be below the $J/\psi$ mass ($M^{\rm
  rec}_{\pi^+\pi^-} < 3.05$ GeV/$c^2$). Background events from $\psip
\ra \eta X$ with $\eta \ra \gamma \pi^+\pi^-$ are greatly suppressed
by the $\eta$ veto, which requires the invariant mass of $\gamma
\pi^+\pi^-$ be outside the $\eta$ signal region,
$M_{\gamma\pi^+\pi^-}>0.554$ or $M_{\gamma\pi^+\pi^-}<0.538$
GeV/$c^2$.

The $\chi^2_{\rm 4C}$ and $\eta$ mass window requirements are
optimized by maximizing $S/\sqrt{S+B}$, where $S$ and $B$ are the
numbers of expected $\etacp$ signal and background events in the
$\etacp$ signal region in data, respectively. The $\etacp$ signal
region is defined as (3.60, 3.66) GeV/$c^2$.  $S$ is calculated by
$S=N_{\psip}^{\rm tot} \BF[\psip \ra \gamma \etacp] \BF[\etacp \ra
  \tripipi] \epsilon$, where $N_{\psip}^{\rm tot}$ is the number of
$\psip$ events~\cite{psipNum} and $\epsilon$ is the detection
efficiency. Without using experimental information on $\BF[\etacp \ra \tripipi]$, 
we assume $\BF[\etacp \ra \tripipi]/\BF[\etacp \ra K_S^0K^-\pi^+\pi^+\pi^-] = \BF[\eta_c(1S)
  \ra \tripipi]/\BF[\eta_c(1S) \ra K_S^0K^-\pi^+\pi^+\pi^-]$. Here
$\BF[\eta_c(1S) \ra \tripipi]$, $\BF[\eta_c(1S, 2S) \ra
  K_S^0K^-\pi^+\pi^+\pi^-]$, and $\BF[\psip \ra \gamma\etacp]$ are
taken from the world average values~\cite{pdg}, and $B$ is estimated
with the generic MC sample.

\section{Background estimation}

The analysis of the generic MC sample for the $\psip$ decays with TopoAna~\cite{topology} indicates that the
dominant background contributions come from two sources: (1) $\psip
\ra \tripipi$ with a fake photon or a photon from FSR in
the final state; (2) $\psip \ra \pi^0\tripipi$ with $\pi^0$
decaying into a $\gamma\gamma$ pair. The remaining background events
are from hundreds of decay modes with small contributions to the
signal processes, which distribute smoothly in the $\tripipi$ invariant
mass spectrum.  In the $\chi_{cJ}$ signal region from 3.35 to 3.58
GeV/$c^2$, the possible peaking background processes, $\chi_{cJ} \ra
K_S^0K3\pi$, $K^+K^-2(\pi^+\pi^-)$,
$K_S^0K_S^0\pi^+\pi^-$, and $\mu^+\mu^-2(\pi^+\pi^-)$, are studied,
and the contaminations from these processes are found to be negligible.

\subsection{\boldmath $\psip \ra \tripipi$}
\begin{figure}[h!]
             \centering
             \includegraphics[width=0.45\textwidth]{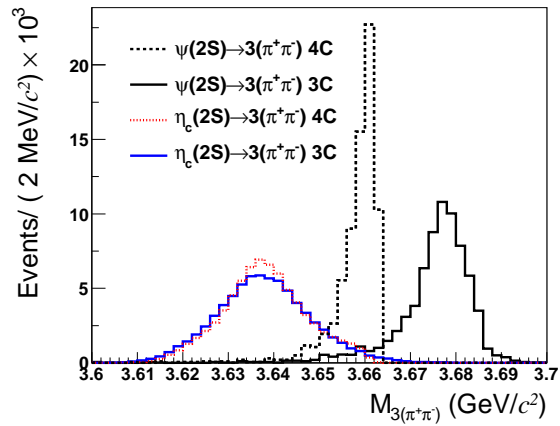}
             \caption{Comparison between 3C and 4C kinematic fits (unnormalized). Shown in the plot are the invariant mass distributions of $\tripipi$ of signal events from the 3C fit (blue solid line) and 4C fit (red dashed line), those of  background events of $\psip \ra \tripipi$ from the 3C fit (black solid line) and 4C fit (black dashed line).}
             \label{3c-4c-comp}
     \end{figure}

The background from $\psip \ra \tripipi$ with a fake photon satisfying
the 4C fit contributes to a peak near the $\etacp$ mass and decreases
rapidly with higher mass in the $\tripipi$ mass spectrum due to the
threshold of 25 MeV for a photon. The inclusion of a fake photon in
the 4C kinematic fit shifts the $\tripipi$ invariant mass peak lower
compared to the true mass. This shift can be corrected by performing a
modified kinematic fit in which the energy of the measured photon is
allowed to vary in the fit (3C fit).  The $\eta_c(2S)$ mass resolution
from the 3C fit is similar to that from the 4C fit, while the former
can separate the background events from the $\eta_c(2S)$ signal
significantly (see Fig.~\ref{3c-4c-comp}).  Therefore, the invariant
mass spectrum from the 3C fit, $M^{\rm 3C}_{\tripipi}$, is used to
determine the signal yield.

The background events from the $\psip \ra \tripipi$ process with an FSR
photon have the same final state as the signal process and can
contaminate the $\etacp$ signal due to a long tail from 3.60 to 3.66
GeV/$c^2$ in the $M^{\rm 3C}_{\tripipi}$ distribution. The size of the
tail depends on the FSR fraction, which is defined as
$R_{\rm FSR}=N_{\rm FSR}/N_{\rm nonFSR}$, where $N_{\rm FSR}$ ($N_{\rm
  nonFSR}$) is the number of events with (without) an FSR photon
surviving the selection. The difference in the FSR fraction between
data and MC simulation is studied by using a control sample $\psip \ra
\gamma \chi_{c0},~\chi_{c0} \ra (\gamma_{\rm FSR})\tripipi$ and
subsequently corrected for by the FSR correction factor.

\begin{figure}[h!]
             \centering
             \includegraphics[width=0.5\textwidth]{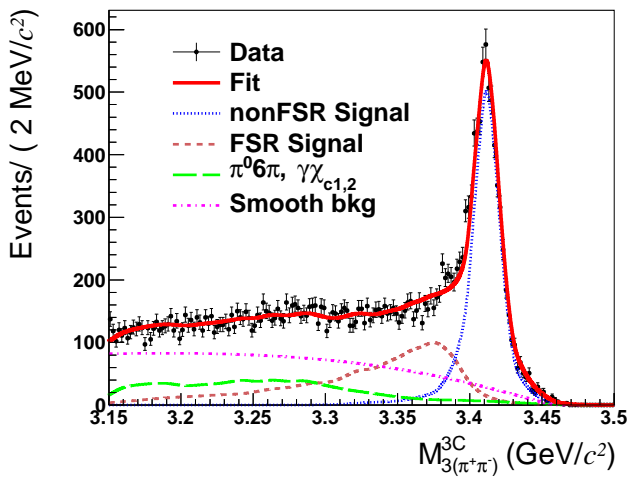}
             \caption{Fitted distribution of $M_{\tripipi}^{\rm 3C}$ from $\psip \ra \gamma \chi_{c0}, ~\chi_{c0} \ra (\gamma_{\rm FSR})\tripipi$ events. The black dots with error bars are the data, the red solid line is the best fit result, the blue dotted line represents the nonFSR component, the brown dashed line represents the FSR component, the green long-dashed line shows the contribution from known background components, and the pink dash-dotted line represents the smooth shape used to describe the remaining background events.}
             \label{fsr-fit}
     \end{figure}

The FSR photon in the control sample is required to be reconstructed. 
The event selection criteria of the
control sample are similar to those of the signal sample, except that
the final states contain two photons. The softer photon is treated as
the FSR photon, and in the 3C fit, the energy of the FSR photon is
free. 
The process $\psip \ra \pi^0\tripipi$ is a dominant background contribution to the control sample, since it has the same final state particles as the control sample. The candidate events satisfying $0.115<M_{\gamma\gamma}<0.150$ GeV/$c^2$ are rejected to remove the background events from $\psip \ra \pi^0\tripipi$, where $M_{\gamma\gamma}$ is the invariant mass of $\gamma\gamma$.  The requirement of $E_{\rm hard}<0.2$ GeV is placed to remove the background events from $\psip \ra \gamma \chi_{c1,2}, \chi_{c1,2} \ra (\gamma_{\rm FSR})\tripipi$, where $E_{\rm hard}$ is the energy of the harder photon.
For this sample, $R^{\rm Data}_{\rm FSR}$ is determined by fitting the
$M^{\rm 3C}_{\tripipi}$ spectrum.  The FSR and nonFSR events are
described by the corresponding MC shapes determined from the $\psip
\ra (\gamma_{\rm FSR})\tripipi$ MC simulation and convolved with a
Gaussian function to account for the resolution difference between
data and MC simulation. The parameters of the Gaussian function are
free in the fit.  The background events from $\psip \ra \pi^0\tripipi$
and $\psip \ra \gamma \chi_{c1,2}, \chi_{c1,2} \ra (\gamma_{\rm
  FSR})\tripipi$ are dominant, and their shapes are determined
directly from the MC simulated events surviving the
$\gamma\gamma\tripipi$ selection. The distribution of the remaining
background events is smooth and thus modeled by an ARGUS
function~\cite{argus}. The number of events of each component is a
free parameter, and the fit result is shown in Fig.~\ref{fsr-fit}.
From the fit, we obtain $R_{\rm FSR}^{\rm Data}=0.70\pm0.05$.  From
the MC simulation, we obtain $R_{\rm FSR}^{\rm MC} = 0.43$. The FSR
correction factor is defined as $f_{\rm FSR} =R_{\rm FSR}^{\rm Data}/R_{\rm FSR}^{\rm MC}= 1.62\pm0.13$, where the uncertainty is statistical. 

Though emitting an FSR photon changes the charge conjugate parity of the $\tripipi$ system, the fact that the $f_{\rm FSR}$ only accounts for the difference on FSR fraction between data and MC simulation and the FSR fraction in MC simulation depends on the masses of the mother and daughter particles~\cite{photos} proves the independence of the $f_{\rm FSR}$ on the charge conjugate parity of the system. Thus, the $f_{\rm FSR}$ obtained from $\chi_{c0}$ decay is applied for $\psip$ decay in this work.
In the fit to determine the numbers of $\etacp$ and
$\chi_{cJ}$ signal events, the background shape from $\psip \ra
(\gamma_{\rm FSR})\tripipi$ is described by the sum of MC simulated
shapes $\psip \ra \tripipi$ and $\psip \ra \gamma_{\rm FSR}\tripipi$
with the FSR fraction corrected by $f_{\rm FSR}$.

\subsection{\boldmath  $\psip \ra \pi^0\tripipi$}
The background from $\psip \ra \pi^0\tripipi$ is measured from data by
reconstructing the $\gamma\gamma\tripipi$ events, where the
$\gamma\gamma$ pair forms a $\pi^0$ candidate. In the case of more
than two photons, the $\pi^0$ candidate with the minimum $\chi^2$ from
a 5C kinematic fit (4C fit plus a $\pi^0$ mass constraint) is selected. The
background contribution is suppressed with the requirement
$\chi^2_{\rm 5C}<50$. The other selection criteria are the same as in
the signal selection. Imposing the signal selection, the background shape of
$\psip \ra \pi^0\tripipi$ is estimated via
     \begin{equation} 
          \nonumber
                \left(\frac{dN}{dM_{\tripipi}}\right)^{\rm 3C}  = \left(\frac{dN}{dM_{\tripipi}}\right)^{\rm 5C}\times \frac{\epsilon^{\rm 3C}_{\gamma\tripipi}}{\epsilon^{\rm 5C}_{\pi^0\tripipi}}, 
     \end{equation}
where $M^{\rm 5C}_{\tripipi}$ is the invariant mass distribution of
$\tripipi$ from the 5C fit for the events from data passing the
$\pi^0\tripipi$ selection, and $\epsilon^{\rm 3C}_{\gamma\tripipi}$
and $\epsilon^{\rm 5C}_{\pi^0\tripipi}$ are the efficiencies as
functions of the $3(\pi^+ \pi^-)$ mass with
which the $\psip \ra \pi^0\tripipi$ MC simulated events pass the
$\gamma\tripipi$ and $\pi^0\tripipi$ selections, respectively.

\subsection{Continuum contribution}
The background contribution from the continuum processes (including the
initial state radiation) is estimated using the datasets taken at the
c.m.\ energies ($E_{\rm c.m.}$) of 3.581 and 3.670 GeV. The momenta and
energies of final state particles are scaled to account for the
difference in c.m.\ energy. The mass spectrum is normalized based on
the differences in the integrated luminosity and cross section. The
energy dependence of the cross section is measured by
the {\it BABAR} experiment~\cite{cross-section}, and we determine it
to be proportional to $1/s^{2.21\pm0.67}$ ($s=E^2_{\rm c.m.}$) by a
fit to the measured cross
sections of $e^+e^- \ra 3(\pi^+\pi^-)$, where the uncertainty includes
the statistical and systematic uncertainties.

 \section{Signal determination} 

\begin{figure}[h!]
	\begin{center}
		\includegraphics[width=0.48\textwidth]{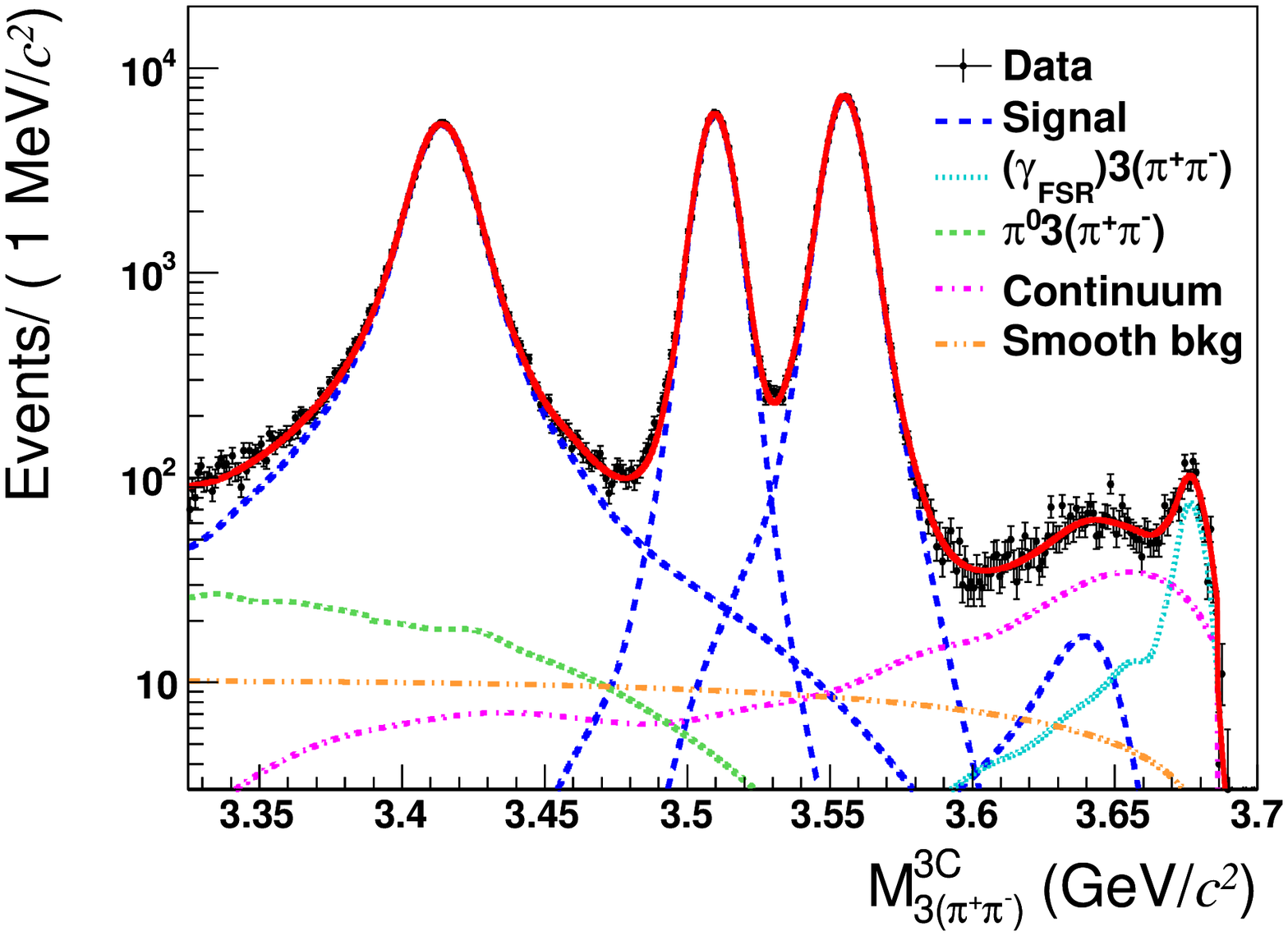}
        \quad
		\includegraphics[width=0.48\textwidth]{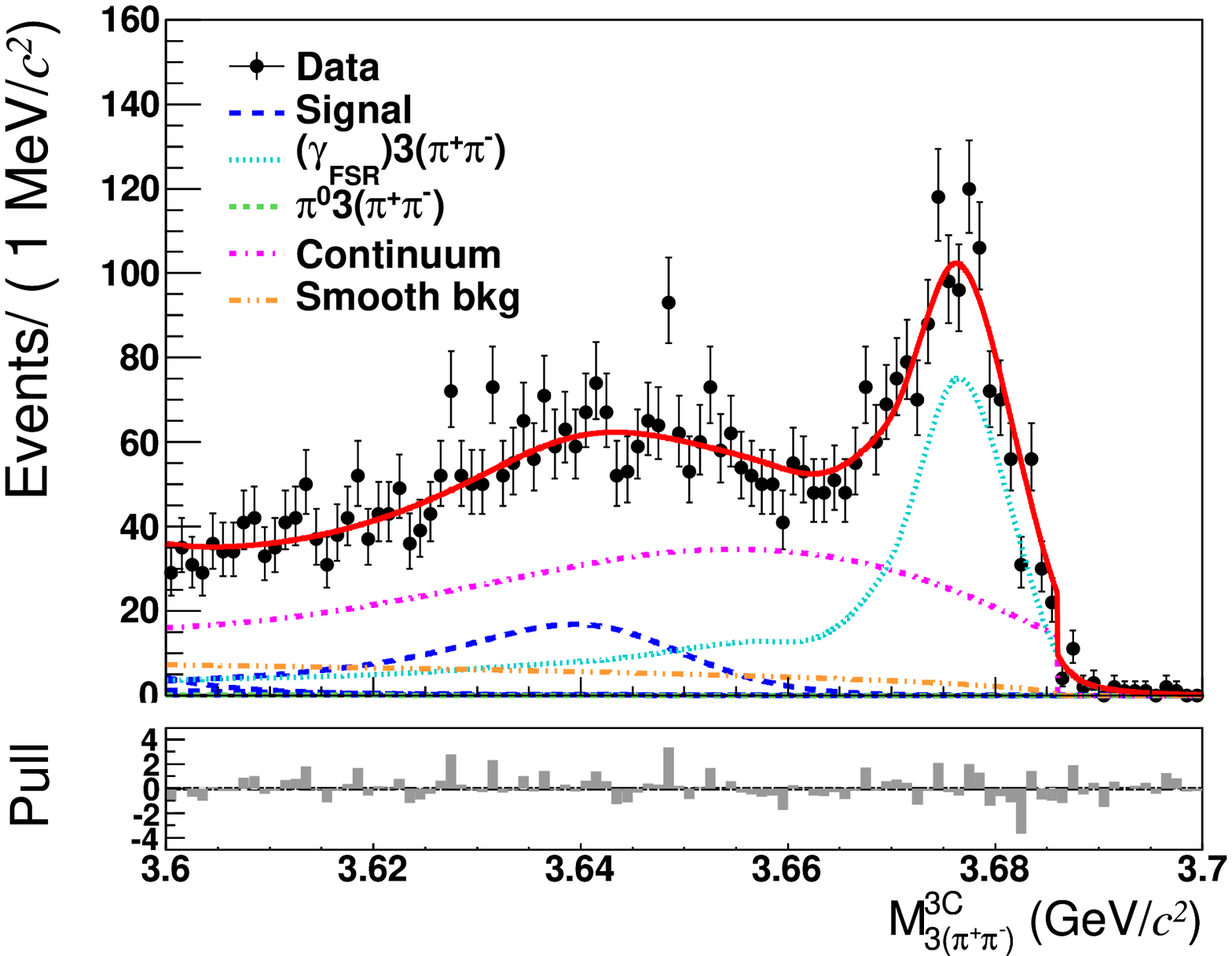}  
		\caption{Invariant mass distributions of $\tripipi$
		after a 3C fit in the whole fit range (top) and in
		the range only containing $\etacp$ signal
		(bottom). Dots with error bars are data, the red solid
		curve is the best fit result, the blue long-dashed
		lines show the $\etacp$ and $\chi_{cJ}$ signal shapes,
		the cyan dotted line represents the contribution from
		$\psip \ra (\gamma_{\rm FSR})\tripipi$, the green
		dashed line shows the contribution from $\psip \ra
		\pi^0\tripipi$, the pink dash-dotted line is the
		continuum contribution, and the orange
		dash-dot-dotted line represents the smooth
		background. }
		\label{final-fit}
	\end{center}
\end{figure}

 The signal yields are determined by a fit to the $M^{\rm
   3C}_{\tripipi}$ spectrum using an unbinned maximum likelihood
 method, as shown in Fig.~\ref{final-fit}. The fit range is from
 3.325 to 3.700 GeV/$c^2$, which includes the $\chi_{cJ}$ signals. The
 line shape of $\etacp$ is described by
\begin{equation} 
          \nonumber
          [E_{\gamma}^3 \times BW(m) \times f_d(E_{\gamma})\times \epsilon(m)] \otimes G(\delta m,\sigma), 
\end{equation} 
where $m$ is the mass of $\tripipi$, $E_{\gamma}$ is the energy of the
transition photon in the rest frame of $\psip$, $BW(m)$ is the
Breit-Wigner function, $f_d(E_{\gamma}$) is a function to damp the
diverging tail from $E_{\gamma}^3$, $\epsilon(m)$ is the efficiency
curve as a function of $m$, and $G(\delta m,\sigma)$ is a Gaussian
function describing the detector resolution determined from MC
simulation. The $f_d(E_{\gamma}$) form proposed by the KEDR
Collaboration~\cite{kedr}, $E_0^2/[E_{\gamma}E_0 +
  (E_{\gamma}-E_0)^2]$, is used in the nominal fit, where $E_0$ is the
most probable energy of the transition photon. The efficiency curve is
parametrized by $(m/{\rm GeV}/c^2)(1-(m/3.6747 ~{\rm
  GeV}/c^2)^2)^{0.303} \times e^{-3.75(1-(m/3.6747~{\rm GeV}/c^2)^2)}$,
obtained by fitting the efficiencies determined at each $m$ using an
ARGUS function~\cite{argus}. The $\chi_{cJ}$ line shapes are obtained
directly from the MC simulations. The $\etacp$ and $\chi_{cJ}$ line shapes
are convolved with additional Gaussian functions to account
for the mass resolution difference between data and MC
simulations. For $\chi_{cJ}$, the parameters of the Gaussian function
are determined in the fit, while for $\etacp$ they are fixed to the values
extrapolated from the $\chi_{cJ}$ results assuming a linear
relationship. The background shapes have been estimated as described above. For
$\psip \ra \pi^0\tripipi$ and the continuum contributions, the shapes
and the numbers of events are fixed. 
For $\psip \ra (\gamma_{\rm FSR})\tripipi$, the shape is from MC simulation convolved with a
Gaussian function with floating parameters and the number of events
is left free. In MC simulation, the fraction of FSR events is corrected by $f_{\rm FSR}$. The remaining background distribution is smooth, thus is
described by an ARGUS function with the number of events a free
parameter in the fit. It is assumed there is no interference between the signal and
the continuum events. With toy MC samples, we validate that the output
values of the mass and width of the $\etacp$ signal, as well as the
numbers of $\etacp$ and $\chi_{cJ}$ signal events, are consistent with
the inputs within one standard deviation, suggesting that the
fit procedure has no bias.

The signal yields ($N^{\rm sig}_{\rm data}$) obtained from the fit
are summarized in Table~\ref{tab:fitresult}.  The $\chi^2/\rm ndf$
value of the fit is $557.2/359 = 1.51$, where ndf is the number of
degrees of freedom. The statistical significance of $\etacp$ is
calculated to be 10.8$\sigma$ from the difference of the logarithmic
likelihoods~\cite{significance}, $-2 \ln(
\mathcal{L}_0/\mathcal{L}_{\rm max})$, where $\mathcal{L}_{\rm
  max}$ and $\mathcal{L}_0$ are the maximized likelihoods with and
without the signal component, respectively. The difference in the
number of degrees of freedom ($\Delta \rm ndf$=3) has been taken into
account. The largest systematic uncertainty, as described in
Sec.~\ref{sec:syserr}, is from the fit to the mass
spectrum. Alternative fits to the $M_{\tripipi}^{\rm 3C}$ spectrum under
different fit conditions are performed, and the $\etacp$ signal
significance is larger than 9.3$\sigma$ in all cases. We measure the
mass and width of $\etacp$, which are $M=(3643.4 \pm 2.3 \pm 4.4)$
MeV/$c^2$ and $\Gamma = (19.8 \pm 3.9 \pm 3.1)$ MeV, respectively. The
branching fraction is calculated using
\begin{equation}
          \nonumber
          \BF[X \ra \tripipi] = \frac{N^{\rm sig}_{\rm data}}{N_{\psip}^{\rm tot}\times \BF(\psip \ra \gamma X)\times \epsilon_{\rm corr}}, 
\end{equation}
where $X$ refers to $\etacp$ or $\chi_{cJ}$, $\BF(\psip \ra \gamma X)$
is the branching fraction of $\psip \ra \gamma X$~\cite{pdg}, and
$\epsilon_{\rm corr}=\epsilon_{\rm MC}*f_{\rm corr}$ is the corrected
signal detection efficiency. The correction factor of the efficiency
takes into account small differences between data and simulation in
the single-track reconstruction efficiency; it
is $f_{\rm corr} = [\sum_{\rm i}^{N_{\rm sel}}\prod_{\rm j}^6
  w_{ij}(p_t, \cos\theta)]/N_{\rm sel}$, where $N_{\rm sel}$ is the
number of signal MC simulated PHSP events surviving the event
selection, $i$ and $j$ run over the surviving events and the six charged
tracks, respectively, $w_{ij}$ is the charged track reconstruction
weight factor in bins of $(p_t, \cos\theta)$, and $p_t$ is the
transverse momentum of the track. The values of $w_{ij}$ are obtained
using the control sample $J/\psi \ra \pi^+\pi^-\pi^0$.  $f_{\rm corr}$
is calculated using a sampling method, where $w_{ij}$ is sampled
according to $G(w_{ij}, \Delta w_{ij})$. Here $\Delta w_{ij}$ is the
uncertainty of $w_{ij}$, $G$ is a Gaussian function. Using 10000
samples, the resulting $f_{\rm corr}$ follows a Gaussian
distribution. The mean value of the Gaussian distribution is taken as
the nominal efficiency correction factor, and the standard deviation
is taken as the uncertainty, labeled as $\Delta f_{\rm
  corr}$. Table~\ref{tab:fitresult} lists the corrected efficiencies
and calculated branching fractions, where the first uncertainty is
statistical, the second systematic, and the third from the uncertainty of
$\BF[\psip\ra\gamma\etacp]$.

Our measurement indicates that the branching fractions for $\chi_{cJ}
\ra \tripipi$ are about twice as large as the current world average.
To validate our results, we perform a number of cross-checks.
We divide the dataset into two
sub-datasets, collected in 2009 and 2012, respectively, and compare
the branching fractions of $\chi_{cJ} \ra \tripipi$ obtained from the two
sub-datasets. The results are consistent with each other and agree
with the nominal results. Additionally, two individual studies of
measurements of branching fractions of $\chi_{cJ} \ra \tripipi$ are
performed and the results from two cross-checks agree well with the
reported results.

    \begin{table*}[htbp]
      \caption{\label{tab:fitresult} Signal yields, corrected signal
        efficiencies, and branching fractions for $\etacp$ and
        $\chi_{cJ}$ decays. The branching fractions of $\chi_{cJ} \ra
        \tripipi$ from the world average values~\cite{pdg} are also
        shown.}
	\vspace{0.2cm}
	\setlength{\tabcolsep}{5mm}{
	\begin{tabular}{c@{$\ra$}c c@{$\,\pm\,$}c c@{$\,\pm\,$}c c@{$\,\pm\,$}c@{$\,\pm\,$}l c}
	\hline\hline
	\multicolumn{2}{c}{Channel}  & \multicolumn{2}{c}{$N^{\rm sig}_{\rm data}$} &\multicolumn{2}{c}{$\epsilon^{\rm corr}$ (\%)}  &\multicolumn{3}{c}{$\BF_{\rm measured}$ ($\times 10^{-2}$)}  &$\BF_{\rm PDG}$ ($\times 10^{-2}$) \\ \hline 
    $\etacp$    &$\tripipi$    &$568.8$  &$63.3$  &$13.84$ &0.01 &$1.31$ &$0.15$ &$0.17^{+0.64}_{-0.47}$  & -   \\ 
    $\chi_{c0}$ &$\tripipi$    &$145300$ &$396$   &$15.92$ &0.01 &$2.080$ &$0.006$ &$0.068$ & $1.20\pm0.18$   \\ 
    $\chi_{c1}$ &$\tripipi$    &$84317$  &$299$   &$17.67$ &0.01 &$1.092$ &$0.004$ &$0.035$  & $0.54\pm0.14$  \\ 
    $\chi_{c2}$ &$\tripipi$    &$112510$ &$347$   &$16.85$ &0.01 &$1.565$ &$0.005$ &$0.048$  & $0.84\pm0.18$  \\ \hline \hline
	\end{tabular} }
\end{table*} 

\section{\label{sec:syserr} Systematic Uncertainties}
Table~\ref{tab:syserr} summarizes the sources of systematic
uncertainties in measuring the branching fractions and the parameters
of $\etacp$. They are described in the following.

\begin{table*}[htbp]
      \caption{\label{tab:syserr} Relative systematic uncertainties in
        the measurements of the branching fractions (in \%), where $f
        = \tripipi$, and absolute systematic uncertainties in the
        $\etacp$ mass (MeV/$c^2$) and width (MeV) measurements. For
        the $\etacpD$ mode, the value in parentheses is the total
        systematic uncertainty without including the uncertainty from
        $\BF(\psip \ra \gamma\etacp)$~\cite{bfr6}.}
             \setlength{\tabcolsep}{2.2mm}{
             \begin{tabular}{l c c c c c c} \hline \hline
              Sources         &$\BF(\etacp \ra f)$ &$\BF(\chi_{c0} \ra f)$ &$\BF(\chi_{c1} \ra f)$ &$\BF(\chi_{c2} \ra f)$  & Mass  & Width    \\ \hline 
              Efficiency correction factor    & 0.07    & 0.06   & 0.06   & 0.06  & -  & -  \\
              Photon reconstruction  & 1.00    & 1.00   & 1.00   & 1.00  & -  & -  \\ 
              Kinematic fit          & 0.70    & 0.53   & 0.53   & 0.74  & -  & -  \\ 
              $J/\psi$ veto          & 2.36    & 0.81   & 0.92   & 1.01  & 0.90  & 1.08  \\  
              $\eta$ veto            & 1.55    & 0.50   & 0.62   & 0.74  & 0.60  & 0.48  \\ 
              Damping function form  & 1.20    & 0.10   & 0.05   & 0.00  & 1.80         & 1.19  \\ 
              Efficiency curve       & 0.63    & 0.08   & 0.01   & 0.10  & 0.80         & 0.88   \\ 
              Gaussian resolutions   & 3.36  & 0.01   & 0.01   & 0.01  & 1.34  & 0.66  \\ 
              Size of FSR correction factor   & 2.23    & 0.05   & 0.01   & 0.01  & 0.20         & 0.28   \\ 
              Number of $\psip \ra \pi^0 \tripipi$ events  & 0.12  & 0.07   & 0.01   & 0.01 & 1.40         & 0.09 \\  
              Shape of $\psip \ra \pi^0 \tripipi$   & 0.07  & 0.18   & 0.01   & 0.01 & 0.60         & 0.04 \\ 
              Number of continuum events   & 7.04    & 0.12   & 0.09   & 0.05   & 1.34         & 1.82  \\ 
              Shape of continuum     & 4.52  & 0.04   & 0.01   & 0.02  & 2.40         & 0.95  \\ 
              MC simulation         & 1.43    & 1.88   & 1.16   & 1.13  & -  & -  \\  
              Possible interference  & 8.61   & -      & -      & -     & 1.60  & 1.05  \\ 
              Number of $\psip$ events     & 0.60    & 0.60   & 0.60   & 0.60  & -  & -  \\ 
              Branching fraction     & $^{+48.57}_{-35.71}$    & 2.04   & 2.46   & 2.10  & -  & -  \\ 
              MC statistics           & 0.34    & 0.58   & 0.54   & 0.56  & -  & -  \\ 
              Total                  & $^{+50.33}_{-38.07}$ (13.20)     & 3.26   & 3.25   & 3.08  & 4.40  & 3.07  \\ \hline \hline
              \end{tabular}  
              }
\end{table*} 
      
The systematic uncertainty from the efficiency correction factor is
given by $\Delta f_{\rm corr}/f_{\rm corr}$.

Based on the study of the photon detection efficiency using the control
samples of $J/\psi \ra \rho^0 \pi^0$ and $e^+e^- \ra \gamma
\gamma$~\cite{photonRec}, the systematic uncertainty due to photon
reconstruction is 1\% per photon.

In the kinematic fit, the helix parameters of charged tracks in MC
simulation have been corrected to improve the consistency between data
and simulation~\cite{helixfit}. The systematic uncertainty from
the kinematic fit is taken as half of the difference between the
efficiencies with and without track helix parameter correction.

To estimate the uncertainty from the $J/\psi$ veto, we vary the maximum
requirement on $M^{\rm rec}_{\pi^+\pi^-}$ from 3.00 to 3.08 GeV/$c^2$
in steps of 0.01 GeV/$c^2$. The maximum difference of the results with
respect to the nominal one is taken as the systematic uncertainty from
the $J/\psi$ veto.  To estimate the uncertainty from the $\eta$ veto,
the mass window of the $\eta$ veto is changed to be (0.538, 0.552),
(0.536, 0.554), (0.534, 0.556), (0.532, 0.558), (0.530, 0.560),
(0.528, 0.562), or (0.526, 0.564) GeV/$c^2$. The results under the
different windows are obtained, and the maximum difference with the
nominal one is taken as the systematic uncertainty from the $\eta$ veto.

An alternative damping function used by the CLEO
Collaboration~\cite{cleo}, $f_d(E_{\gamma})=
\exp(-E_{\gamma}^2/8\beta^2)$, is chosen to estimate the uncertainty
from the damping function, where $\beta$ is a free parameter. The
difference between the two damping functions is taken as the
systematic uncertainty.  We model the efficiency curve using a
threshold function with the parametrized form of
$\sqrt[4]{3.675-(m/{\rm
    GeV}/c^2)}\left[1.5-1.5(3.675-(m/(\rm{GeV}/c^2))\right]$, and take
  the difference between the two functions as the systematic
  uncertainty.  The resolutions of two Gaussian functions for the line
  shape of the $\etacp$ signal are varied by $\pm1\sigma$, and the
  largest difference is taken as the systematic uncertainty.

The systematic uncertainties related to the background contributions
are from the FSR process, continuum, and $\psip \ra \pi^0\tripipi$, of
which the first two may influence our fit results in the $\etacp$ mass
region. We vary the FSR correction factor by $\pm1\sigma$ and take the
largest difference as the uncertainty from the shape of FSR process.
The number of continuum events is determined by the normalization factor
calculated using the integrated luminosity and cross section, where
the latter is proportional to $1/s^{2.21\pm0.67}$. Taking into account
a 1.0\% uncertainty of the integrated luminosity, a new normalization
factor is calculated, and the mass spectrum is fitted with the new
fixed number of continuum events. The differences of the measured
results with respect to the nominal ones are taken as the uncertainty
from the luminosity. Additionally, we change the cross section by
varying the exponent by $\pm0.67$ and refit the mass spectrum. The
uncertainty from the cross section is estimated to be the difference
of the measured result.  Finally, the systematic uncertainty from
the number of continuum events is determined by adding the uncertainties from
the luminosity and the cross section in quadrature.  We modify the
number of the background events of $\psip \ra \pi^0\tripipi$ by
$\pm1\sigma$ to estimate the uncertainty from the number of $\psip \ra
\pi^0\tripipi$ events. The line shapes of continuum and $\psip
\ra \pi^0\tripipi$ background are smoothed by
RooKeysPDF~\cite{keyspdf} in the nominal fit.  We use
RooHistPDF~\cite{histpdf} and take the difference under the two methods as
the systematic uncertainty.
     
\begin{figure}[h!]
	 \begin{center}
		\includegraphics[width=0.45\textwidth]{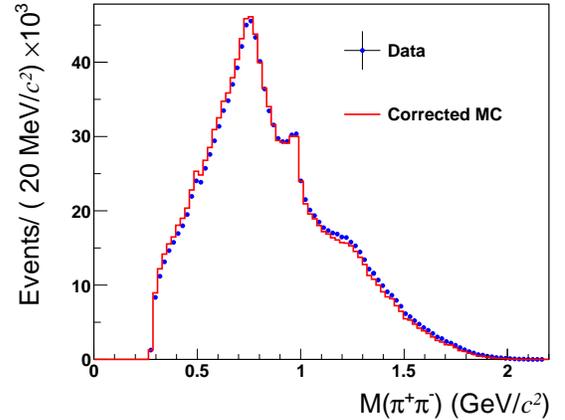}
		\caption{Invariant mass distributions of $\pi^+\pi^-$ from data and the corrected MC simulation. The blue dots represent the data and the red solid line stands for the corrected MC simulation. }
		\label{fig:Mpipi}
	 \end{center}
\end{figure}

There are obvious $\rho$ and $f_0(980)$ intermediate states in the
$M_{\pi^+\pi^-}$ distribution from the data, as shown in
Fig.~\ref{fig:Mpipi}, while these states are not taken into account in
the PHSP MC simulation. We correct the $M_{\pi^+\pi^-}$ distribution
from PHSP MC simulation to agree with data, as shown by the red
line in Fig.~\ref{fig:Mpipi}. Additionally, the distributions of
track momenta disagree between data and MC simulation, and we correct
the MC sample similarly. The difference in detection
efficiency with and without these corrections is taken as the
systematic uncertainty. The uncertainty of assuming $\psip \ra \gamma
\chi_{cJ}$ as a pure electric dipole transition is studied by
considering the contribution from higher-order multipole
amplitudes~\cite{m2e31,m2e32,m2e33} in the MC simulation, and the
differences of the efficiency, 1.0\% for $\chi_{c1}$ and 0.2\% for
$\chi_{c2}$, are taken as the systematic uncertainties. As a
conservative estimate, we assign a 1.0\% uncertainty for assuming
$\psip \ra \gamma \etacp$ as a pure magnetic dipole transition due to
the absence of experimental measurements of higher-order multipole
amplitudes. The systematic uncertainty from the MC simulation is
calculated by adding above uncertainties in quadrature.

There could be interference between the signal process and the
$e^+e^- \ra \gamma\tripipi$ process when the quantum numbers of the two
systems are the same.  To
estimate the uncertainty of the assumption of no interference in the nominal
fit, we include the contribution from possible interferences into
the mass spectrum fit and take the maximum differences of the measured
branching fractions and the measured mass and width of $\etacp$
relative to the nominal ones as the uncertainty.  Using the difference
of the maximum likelihoods and the number of degrees of
freedom with and without the interference effect, the significance of the interference contribution is calculated to be 0.54$\sigma$. 

The number of $\psip$ events is determined to be
$(448.1\pm2.9)\times10^6$~\cite{psipNum}; therefore 0.6\% is taken as
the uncertainty.  The systematic uncertainties from branching
fractions of $\psip \ra \gamma X$ are taken from the world average
values~\cite{pdg}, where $X$ is $\etacp$ or $\chi_{cJ}$. The
uncertainty from the statistics of the signal MC sample is considered as
well.

\section{Conclusion}
Using $(448.1\pm2.9)\times10^{6}$ $\psip$ events collected by the
BESIII detector, the hadronic decay $\etacp \ra 3(\pi^+\pi^-)$ is
observed for the first time with a significance of 9.3$\sigma$.

The measured mass of $\etacp$ is $(3643.4 \pm 2.3 \pm 4.4)$ MeV/$c^2$,
and the width is $(19.8 \pm 3.9 \pm 3.1)$ MeV, which are consistent
with the world average values within $2\sigma$~\cite{pdg}.  The
product branching fraction $\BF[\psip \to \gamma
\eta_c(2S)]\times\BF[\eta_c(2S)\to3(\pi^+\pi^-)]$ is measured to be
$(9.2 \pm 1.0 \pm 1.2 )\times10^{-6}$. Using $\BF[\psip \to \gamma
\eta_c(2S)]=(7.0^{+3.4}_{-2.5})\times10^{-4}$~\cite{bfr6}, we obtain
$\BF[\etacp \ra 3(\pi^+\pi^-)] = (1.31 \pm 0.15 \pm 0.17
~^{+0.64}_{-0.47})\times10^{-2}$, where the first uncertainty is
statistical, the second systematic, and the third from $\BF[\psip \ra \gamma
\etacp]$.

Using $\BF[\eta_c(1S)\ra\tripipi] = (1.7\pm0.4)\%$~\cite{pdg}, we
calculate the ratio of branching fractions
\begin{equation}
     \nonumber
     \frac{\BF[\eta_c(2S) \ra \tripipi]}{\BF[\eta_c(1S)\ra\tripipi]} = 0.77 \pm 0.59,
\end{equation}
where the uncertainty is obtained by assuming the systematic
uncertainties of both branching fractions are uncorrelated. This central value
seems to lean slightly towards the prediction from Ref.~\cite{bfr5} over the
prediction from Ref.~\cite{bfr4}, but our result is compatible with
both predictions.
     
We update the branching fractions of $\chi_{cJ} \ra \tripipi$, which
are summarized in Table~\ref{tab:fitresult}.  Compared to the world
average values~\cite{pdg}, our measured values are almost twice as
large.

\section{ACKNOWLEDGMENTS}
     The BESIII collaboration thanks the staff of BEPCII and the IHEP computing center for their strong support. This work is supported in part by National Key R\&D Program of China under Contracts Nos. 2020YFA0406300, 2020YFA0406400; National Natural Science Foundation of China (NSFC) under Contracts Nos. 11635010, 11735014, 11835012, 11935015, 11935016, 11935018, 11961141012, 12022510, 12025502, 12035009, 12035013, 12192260, 12192261, 12192262, 12192263, 12192264, 12192265; the Chinese Academy of Sciences (CAS) Large-Scale Scientific Facility Program; Joint Large-Scale Scientific Facility Funds of the NSFC and CAS under Contract No. U1832207, U2032108;
     CAS Key Research Program of Frontier Sciences under Contract No. QYZDJ-SSW-SLH040; 100 Talents Program of CAS; The Institute of Nuclear and Particle Physics (INPAC) and Shanghai Key Laboratory for Particle Physics and Cosmology; ERC under Contract No. 758462; European Union's Horizon 2020 research and innovation programme under Marie Sklodowska-Curie grant agreement under Contract No. 894790; German Research Foundation DFG under Contracts Nos. 443159800, Collaborative Research Center CRC 1044, GRK 2149, FOR 2359; Istituto Nazionale di Fisica Nucleare, Italy; Ministry of Development of Turkey under Contract No. DPT2006K-120470; National Science and Technology fund; National Science Research and Innovation Fund (NSRF) via the Program Management Unit for Human Resources \& Institutional Development, Research and Innovation under Contract No. B16F640076; STFC (United Kingdom); Suranaree University of Technology (SUT), Thailand Science Research and Innovation (TSRI), and National Science Research and Innovation Fund (NSRF) under Contract No. 160355; The Royal Society, UK under Contracts Nos. DH140054, DH160214; The Swedish Research Council; U. S. Department of Energy under Contract No. DE-FG02-05ER41374.


%

\end{document}